\documentclass[icps3]{svjour}
\usepackage{latexsym}
\usepackage{graphics}

\begin{document}
\title{Frequency dependent transport in the integer quantum Hall effect}
\titlerunning{Frequency dependent transport in the integer quantum Hall effect}
\author{A. B\"aker \and L. Schweitzer}            
\authorrunning{A. B\"aker \and L. Schweitzer}
\institute{Physikalisch-Technische Bundesanstalt, 
Bundesallee 100, 38116 Braunschweig, Germany}
\maketitle
\begin{abstract}
The frequency dependent transport is investigated for a
two-dimensional disordered system under QHE conditions. 
The real and imaginary parts of the conductivity are calculated numerically
in linear response using a recursive Green function technique.
The energy dependence of $\sigma_{xx}(E,\omega)$ is obtained within 
the lowest Landau band. When the width of the system exceeds the
characteristic length, $L_{\omega}=(\hbar\omega \rho(E))^{-1/2}$, 
the maximum of the real part of $\sigma_{xx}(E,\omega)$ decays with 
frequency almost linearly which is different from the classical Drude 
behaviour.
\end{abstract}

\section{Introduction}
\label{intro}
The integer quantum Hall effect is usually explained by particular
localisation-delocalisation effects that emerge when electrons move 
in disordered two-dimensional systems in the presence of a strong 
perpendicular magnetic field. These phenomena can be investigated in 
more detail if one considers dynamic processes such as frequency
dependent electrical transport. Although electron-elec\-tron interactions 
are commonly neglected, because they are believed to be of minor
importance in the integer QHE, analytical theories that allow the
calculation of the transport coefficients are not available nowadays.
Therefore, numerical investigations are a practical means for computing
the desired quantities.

In previous numerical work, low frequency anomalies and size scaling
of the real part of the dynamic conductivity have been reported 
\cite{GB94,GB96}. 
Recently, we were able to calculate the frequency dependence of the 
real and imaginary part of $\sigma_{xx}(\omega)$ in the lowest Landau 
tail \cite{BS99} which could be compared with approximate analytical 
predictions \cite{VE90}. Also, the 
frequency scaling of $\Re\sigma_{xx}(\omega)$ in the lowest Landau
band has been discovered where a power law $\sim\omega^{\kappa}$ was 
observed with $\kappa=(z\mu)^{-1}\simeq 0.2$ \cite{BS99} which leads to a
critical exponent $\mu\approx 2.35$ \cite{Huc92} with an assumed $z=2$ for
non-interacting electrons. This is in good agreement with the frequency
scaling results of Engel \textit{et al.} \cite{ESKT93} (except that
$z=1$ due to interactions), but at variance with the conclusions of
Ref.~\cite{BMB98}. An interpretation of the conflicting experimental
findings is given in \cite{KMK99}. 
In the present contribution we focus on both the real and imaginary
parts of the frequency dependent longitudinal conductivity 
$\sigma_{xx}(\omega)$ in the centre, and on the energy dependence within 
the lowest Landau band. 

\section{Model and Method}
\label{sec:1}
We consider a disordered two-dimensional electronic system in a strong
magnetic field with $z$-component $B$ described by a tight binding 
Hamiltonian. The magnetic field, which is expressed by a vector potential 
in Landau gauge, $\mathbf{A}=(0,Bx,0)$, is introduced via complex phase 
factors $V_{(x_m,y_m);(x_m,y_m\pm a)}\sim\exp(\pm i2\pi\alpha_B x_m/a)$ 
in the transfer terms for transitions between neighbouring 
lattice sites along the y-direction \cite{BS99,SKM84a}. 
We choose $\alpha_B=eBa^2/h=1/8$
which corresponds to 1/8 flux quantum per lattice cell $a^2$. The
uncorrelated disorder potentials are distributed uniformly between 
$-W/2$ and $W/2$. The conductivity is calculated
within linear response from the Kubo-formula iteratively by a 
generalisation of a recursive Green function technique introduced
earlier \cite{Mac85,SKM85}. For the frequency dependent longitudinal
conductivity at zero temperature we get
\begin{eqnarray}
\lefteqn{\sigma_{xx}(\omega) =
\lim_{\varepsilon \to 0}\lim_{\Omega \to \infty}
\frac{e^2/h}{\Omega \hbar \omega}\,\Bigg[ 
\int\limits_{E_F-\hbar \omega}^{E_F}\hspace{-.2cm}dE \, 
{\rm Tr}\, \Big[(\hbar\eta)^2 x G_{\omega}^+ x G^{-}}
\nonumber  \\
&-&\ \ (\hbar \omega)^2  x G_{\omega}^+ x G^{+}
+ 2 i \varepsilon x^2(G_{\omega}^+-G^{-}) \Big]
\nonumber \\
&-& 
\int\limits_{- \infty}^{E_F-\hbar \omega}\hspace{-.2cm}dE\, {\rm Tr }\Big[
(\hbar \omega)^2 \,\{ x G_{\omega}^+ x G^{+} -x G_{\omega}^- x G^{-} 
\}\Big]\Bigg],
\end{eqnarray}
where $\Omega=L M$ is the area of the system,
$\eta=\omega+2i\varepsilon/\hbar$, and 
$G_{\omega}^{\pm}=(E-H+\hbar\omega\pm i\varepsilon)^{-1}$ 
are the advanced and retarded Green function, respectively.

\section{Results and Discussion}
\subsection{Frequency dependence at filling factor $\nu=1/2$}
\begin{figure}
\resizebox{0.48\textwidth}{!}{%
  \includegraphics{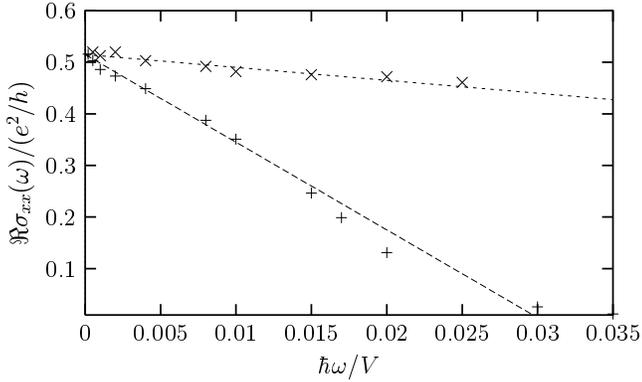}
}
\caption{The real part of the longitudinal conductivity
$\sigma_{xx}(\omega)$ in the centre of the lowest Landau band as a 
function of frequency $\omega$ for disorder strengths $W/V=0.1$ ($+$),
and $W/V=1.0$ ($\times$). The decay of the data, which are shown in 
comparison with linear relations (broken lines), is in contrast to
the classical Drude behaviour. The width and length of the systems are
$M/a=32$ and $L/a=2\cdot 10^6$, respectively.}
\label{nonDrude}      
\end{figure}

The results for the real part of the peak value of the longitudinal 
conductivity $\sigma_{xx}(\omega)$ versus frequency $\omega$ are shown
in Fig.~\ref{nonDrude} for two different values of disorder.
For small frequencies an almost linear decay is observed 
for disorder strengths $W/V=0.1$ and 1.0.
This is in contrast to the classical Drude behaviour given by
\begin{equation}
\sigma_{\textnormal{D}}(\omega)=\sigma_0
\frac{1+\tau^2(\omega_c^2+\omega^2)}%
{1+2\tau^2(\omega_c^2+\omega^2)+\tau^4(\omega_c^2-\omega^2)^2},
\end{equation}
where $\omega_c=eB/m$ is the cyclotron frequency and $\tau$ is the 
collision time. In the limit $\omega\ll\omega_c$, the Drude 
conductivity $\sigma_{\textnormal{D}}(\omega)$ is almost constant 
whereas the peak value of $\Re\sigma_{xx}(\omega)$ clearly 
decays with frequency. A similar
non-Drude decay has been reported in Ref.~\cite{GB96} for a different
type of model with strong correlated disorder potentials. Owing to the
knowledge of the results for a semiclassical model \cite{EB94} the 
deviations from Drude-like behaviour have been attributed \cite{GB96} 
to long time tails
in the velocity correlations which were shown to exist in a QHE system 
in the absence of electron-electron interactions. 
The effects of interactions have been taken into account in
Ref.~\cite{BGK97}.

\begin{figure}[b]
\resizebox{0.48\textwidth}{!}{%
  \includegraphics{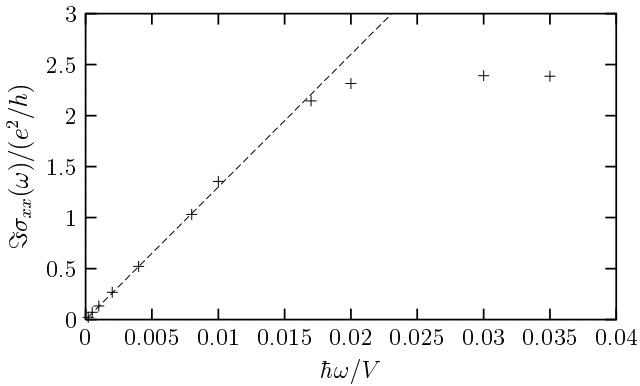}
}
\caption{The imaginary part of the longitudinal conductivity
$\sigma_{xx}(\omega)$ in the centre of the lowest Landau band as a 
function of frequency $\omega$ for disorder strength $W/V=0.1$. 
The width and length of the system is $M/a=32$ and $L/a=2\cdot
10^6$.}
\label{Imsigxx}       
\end{figure}

For the tight-binding model with uncorrelated disorder potentials
considered in the present paper it is not clear whether the picture 
of electron motion along equipotential lines, a basic ingredient
for the arguments of Ref.~\cite{GB96}, is appropriate. Nevertheless,
similar non-Dru\-de behaviour is seen in the frequency dependence of the 
peak value of the real part of $\sigma_{xx}(\omega)$ in the lowest Landau band.
We notice that the frequency decay of $\sigma_{xx}(\omega)$ is
accompanied by the decrease of the characteristic length
$L_{\omega}=(\hbar\omega \rho(E))^{-1/2}$ below the system width $M$. 
Recently, an increase of $\Re\sigma_{xx}(\omega)$ with enhancing
frequency at the QHE transition has been reported for a model based on
Dirac fermions \cite{JZ99}.

The frequency dependence of the imaginary part of $\sigma_{xx}(\omega)$ 
for filling factor near $\nu=1/2$ is shown in Fig.~\ref{Imsigxx}. As
before, the disorder strength and system width are $W/V=0.1$ and $M/a=32$,
respectively. For small frequencies a behaviour
$\Im\sigma_{xx}(\omega)\sim\omega$ is observed. 
The linear increase tapers off at about 
$\hbar\omega/V=0.015$ and saturates at a value of 
$\Im\sigma_{xx}(\omega)\simeq 2.39$.

\begin{figure}
\resizebox{0.475\textwidth}{!}{%
  \includegraphics{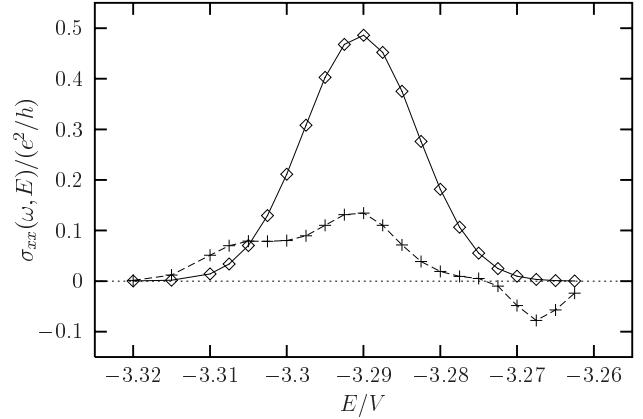}
}
\caption{The energy dependence of the real ($\diamond$) and imaginary 
($+$) part of the
longitudinal conductivity $\sigma_{xx}(\omega,E)$ in the lowest Landau
band. The applied frequency is $\hbar\omega/V=0.001$ and the disorder 
strength $W/V=0.1$. The width and length of the system is $M/a=32$ 
and $L/a=2\cdot 10^6$, respectively.}
\label{endep}     
\end{figure}

\subsection{Energy dependence of $\sigma_{xx}(\omega)$}
The energy dependence of both the real and imaginary parts of
$\sigma_{xx}(\omega)$ are shown in Fig.~\ref{endep} for 
frequency $\hbar\omega/V=0.001$, and disorder strength $W/V=0.1$. 
The system width is $M/a=32$ so that $L_{\omega}$ is smaller than $M$. 
While $\Re\sigma_{xx}(\omega)$ shows the expected conductivity peak the
imaginary part exhibits a weak positive double peak structure before 
it becomes negative at the upper Landau band edge. We note here that the
behaviour of $\Im\sigma_{xx}(\omega)$ changes completely, i.e.,
$\Im\sigma_{xx}(\omega)$ takes negative values for almost any energy, 
if $L_\omega > M$.
Further numerical and analytical work is necessary to elucidate the
microscopic origin of the non-Drude behaviour of the peak
value of $\Re\sigma_{xx}(\omega)$ and the peculiar frequency
dependence of the imaginary part.


\end{document}